%% file: main.tex
\documentclass{tlp}

\usepackage{amsmath}
\usepackage{amssymb} 
\usepackage{wasysym} 
\usepackage{textcomp} 
\usepackage{url}
\usepackage{standalone}
\usepackage[Gray,squaren,thinqspace,thinspace]{SIunits}

\usepackage{spverbatim}
\usepackage{alltt}
\usepackage{xcolor}
\usepackage{soul} 
\usepackage{graphicx}
\usepackage{subfigure}
\usepackage{booktabs}
\newcommand{\TIKZ}[1]{}

\newcommand{\ra}[1]{\renewcommand{\arraystretch}{#1}}

\definecolor{very-light-gray}{gray}{0.95}

\TIKZ{\usepackage{tikz}}
\TIKZ{\usetikzlibrary{external,arrows,trees,fit,positioning,matrix,calc,shapes,intersections,decorations,decorations.pathmorphing,decorations.markings,arrows.meta}}

\newtheorem{exampleTheoremEnv}{Example}

\newenvironment{example}
  {
     \begin{exampleTheoremEnv}
  }
  {
     \end{exampleTheoremEnv}
  }

\newcommand{\xdownarrow}[1]{%
  \ensuremath{%
    \left\downarrow\vbox to #1{}\right.\kern-\nulldelimiterspace
  }%
}

\usepackage{fancyvrb}
\definecolor{frameGrey}{gray}{0.8}
\DefineVerbatimEnvironment{MyFloatingVerbatim}{Verbatim}{fontsize=\footnotesize,frame=single,rulecolor=\color{frameGrey}}
\DefineVerbatimEnvironment{MyNonFloatingVerbatim}{Verbatim}{fontsize=\footnotesize,xleftmargin=1cm}

\title
  [Tabling as a Library]
  {Tabling as a Library with Delimited Control}

\author[Benoit Desouter, Marko van Dooren and Tom Schrijvers]
       {
        BENOIT DESOUTER
        and MARKO VAN DOOREN \\
        Ghent University, Belgium \\
        \email{benoit.desouter,marko.vandooren@ugent.be} \\
        \and
        TOM SCHRIJVERS \\
        KU Leuven, Belgium \\
        \email{tom.schrijvers@cs.kuleuven.be}
       }

\jdate{March 2003}
\pubyear{2003}
\pagerange{\pageref{firstpage}--\pageref{lastpage}}
\doi{the DOI}

\begin{document}

\maketitle

\begin{abstract}
Tabling is probably the most widely studied extension of Prolog. But despite
its importance and practicality, tabling is not implemented by most Prolog
systems. Existing approaches require substantial changes to the Prolog engine,
which is an investment out of reach of most systems. To enable more widespread
adoption, we present a new implementation of tabling in under 600 lines of
Prolog code. Our lightweight approach relies on delimited control and provides
reasonable performance.
\end{abstract}


\section{Introduction}
\input{introduction}
\section{Background: Delimited Continuations}
\label{sec:background}
\input{background}
\section{Shallow Program Transformation}
\input{implementation}

\section{Evaluation}
\input{evaluation}

\section{Related Work}
\input{relatedWork}

\section{Conclusion}
\input{conclusion}

\paragraph{Acknowledgements}
We are grateful to Bart Demoen for supporting hProlog and to the anonymous reviewers for their helpful feedback. This work was partly funded by the Flemish Fund for Scientific Research (FWO).

\bibliographystyle{acmtrans}
\bibliography{main}

\end{document}

%% file: introduction.tex
Tabling is one of the most widely studied extensions to Prolog because it considerably raises the declarative nature of the language.  Tabling
takes away the sensitivity of SLD resolution to rule and goal ordering, and
allows a larger class of programs to terminate. As an added bonus,
the memoisation that is done by the tabling mechanism may drastically improve performance in exchange for more memory.

Given all these advantages, it may come as a surprise that many Prolog systems still do not support tabling. The reason for this is that existing implementations, such as those of Yap and XSB, require pervasive changes to the Prolog engine. This is a substantial engineering
effort that is beyond most systems~\cite{yap}.

Several works have already attempted to tackle this problem. Through the foreign function interface, Ramesh and
Chen~\shortcite{ramesh1994} extend Prolog with tabling primitives implemented in C. A complicated program transformation introduces calls to these C routines at the
appropriate points in tabled predicates. More recently,
Guzm\'an {et.~al.}~\shortcite{guzman2008} have addressed the performance bottlenecks
of Ramesh and Chen's approach. But while their improvement is successful in terms of performance, it does require lower-level
C primitives, changes to the WAM's memory management, and an even more complicated
program transformation. 
These changes further increase the cost of porting and maintaining the mechanism,
and the development effort cannot be amortised over other
features. Hence, the approach does not lower the
threshold for adopting tabling.

Extension tables~\cite{fan1992} provide a tabling mechanism that is implemented directly in Prolog. However, the approach cannot achieve satisfactory performance as suspended goals are always re-evaluated. The initial implementation used the \texttt{assert} and \texttt{retract} predicates for database manipulations. These predicates are notorious for their slow performance. A later version moved the data structures to C, but did not change the inherent recomputation behaviour.

Santos Costa {et~al.}~\shortcite{yap} point out that \emph{``Making it easy to change and control Prolog execution in a flexible way is a fundamental challenge for Prolog.''}.
We argue that delimited control, a language construct for manipulating a program's control flow, does exactly that.
Schrijvers {et.~al.}~\shortcite{iclp2013} show that the impact of delimited control on the WAM
is minimal. On top of that, the development effort of delimited control can be
amortized over the range of high-level language features they enable, such as
\textit{effect handlers}~\cite{plotkinPretnar}.

We show how delimited control can be used for a lightweight tabling mechanism. Both the tabling control flow and
data structures are written entirely in Prolog enhanced with delimited control. It does not require deep custom changes to the Prolog engine,
complicated program transformations, or meta-interpretation. As such our mechanism demystifies many aspects of implementing tabling.



Compared to existing state-of-the-art systems, our system needs more attention in terms of performance, but this does not outweigh the
gain in flexibility: we bring tabling much closer to the masses. In contrast with extension tables, our approach does not require recomputation of suspended goals.  Our tabling implementation is available at \url{http://users.ugent.be/~bdsouter/tabling/}.


%% file: background.tex
Delimited control~\cite{felleisen,danvy} is the key ingredient of our lightweight tabling approach. This technique originates in functional programming and was recently introduced in Prolog by Schrijvers et al.~\shortcite{iclp2013,reset} in the form
of two built-ins: 
\verb!reset/3! and \verb!shift/1! for delimiting and capturing the
continuation respectively.
\begin{itemize}
	\item \texttt{reset(Goal,Cont,Term1)} executes \texttt{Goal}. If \texttt{Goal} calls \texttt{shift(Term2)}, its further execution is suspended and unified with continuation \texttt{Cont}. A continuation is an unspecified Prolog term, which can be resumed using \verb!call/1!. It can be called, saved, copied and compared like any other term, but it is opaque: from its representation we cannot determine anything about the actual goals it represents.

	\item \texttt{shift(Term2)} unifies the remainder of \texttt{Goal}
	up to the nearest call to \texttt{reset/3} (i.e., the delimited continuation) with \texttt{Cont}, and its return value \texttt{Term2} with \texttt{Term1}. Finally, it returns control to just after the \texttt{reset/3} goal.

\end{itemize}

We start with an example that does not call the continuation.
\begin{figure}[ht]
\begin{minipage}{0.30\linewidth}
\begin{Verbatim}[fontsize=\footnotesize]
p :-                
  reset(q,Cont,Term1),
  writeln(Term1),     
  writeln(Cont),      
  writeln(end).      
\end{Verbatim}
\end{minipage}
\quad
\begin{minipage}{0.30\linewidth}
\begin{Verbatim}[fontsize=\footnotesize]
q :-         
  writeln('before shift'),  
  shift('return value'),
  writeln('after shift').  
             
\end{Verbatim}
\end{minipage}
\quad
\begin{minipage}{0.30\linewidth}
\begin{Verbatim}[fontsize=\footnotesize]
?- p.              
before shift
return value              
[$cont$(785488,[])]
end
\end{Verbatim}
\end{minipage}
\end{figure}

This example shows that \verb!shift/1! instantiates the last two arguments
of \verb!reset/3!. \texttt{Cont} represents the \texttt{writeln('after shift')} goal in the
context of the activation of the clause for \texttt{q/0}. But since the continuation is not called, this goal has no effect. \texttt{Term1} is unified with the term \texttt{'return value'}. The execution continues after the \texttt{reset/3}.

The following example shows what happens if the continuation is called:

\begin{figure}[ht]
\begin{minipage}{0.30\linewidth}
\begin{Verbatim}[fontsize=\footnotesize]
p :-                
  reset(q,Cont,Term1),
  writeln(Term1),     
  call(Cont),         
  writeln(end).      
\end{Verbatim}
\end{minipage}
\quad
\begin{minipage}{0.30\linewidth}
\begin{Verbatim}[fontsize=\footnotesize]
q :-         
  writeln('before shift'),  
  shift('return value'),
  writeln('after shift').  

\end{Verbatim}
\end{minipage}
\quad
\begin{minipage}{0.30\linewidth}
\begin{Verbatim}[fontsize=\footnotesize]
?- p.
before shift
return value
after shift    
end 
\end{Verbatim}
\end{minipage}
\end{figure}


%% file: implementation.tex
In our approach, tabled predicates require no special notation, nor any
syntactic ana\-lysis of the predicates being tabled. Predicates are written in the usual way,
and transformed by a shallow program transformation.

\begin{figure}[ht]
\begin{minipage}[t]{0.40\linewidth}
\begin{MyFloatingVerbatim}[fontsize=\footnotesize]
:- table p/2.

p(X,Y) :- p(X,Z), e(Z,Y).
p(X,Y) :- e(X,Y).
\end{MyFloatingVerbatim}
\caption{Running example: transitive closure.\label{fig:runningExample}}
\end{minipage}
\quad
\begin{minipage}[t]{0.50\linewidth}
\begin{MyFloatingVerbatim}[fontsize=\footnotesize]
p(X,Y) :- table(p(X,Y),p_aux(X,Y)).

p_aux(X,Y) :- p(X,Z), e(Z,Y).
p_aux(X,Y) :- e(X,Y).
\end{MyFloatingVerbatim}
\caption{Result of the transformation.\label{fig:transformationResult}}
\end{minipage}
\end{figure}

The use of tabling is illustrated in Figure~\ref{fig:runningExample}. Predicate \texttt{p/2}
computes the transitive closure of the \texttt{e/2} relation.
The \verb!table!-directive indicates that \texttt{p/2} will be
tabled. Predicates without that directive are resolved using standard SLD-resolution.

The \texttt{table/1} directive performs a very shallow program transformation,
the result of which is shown in Figure~\ref{fig:transformationResult}. This
transformation introduces \texttt{p\_aux/2}, which we call the \emph{worker}
predicate, and \texttt{p/2}, the \emph{wrapper} predicate.  The
wrapper predicate is defined in terms of the tabling predicate
\texttt{table/2}, which care of tabling that call fully dynamically. The next section
explains how \texttt{table/2} can be implemented directly in Prolog.

\section{Implementation of the Tabling Library}

This section explains how we implement tabling as a library.

\subsection{The \texttt{table/2} Predicate} \label{subsec:table2predicate}

\begin{figure}
\begin{center}
\begin{minipage}{0.6\textwidth}
\begin{MyFloatingVerbatim}
table(Wrapper,Worker) :-
  get_table_for_variant(Wrapper,Table),
  table_get_status(Table,Status)
  ( Status = complete ->
    get_answer_from_table(Table,Wrapper)
  ;
    ( exists_scheduling_component ->
      run_leader(Wrapper,Worker,Table),
      get_answer_from_table(Table,Wrapper)
    ;
      run_follower(Status,Wrapper,Worker,Table)
    )
  ).
\end{MyFloatingVerbatim}
\end{minipage}
\end{center}
\caption{The \texttt{table/2} predicate.\label{fig:tablePredicate}}
\end{figure}

Thanks to the shallow program transformation, the \texttt{table/2} predicate
intercepts every call to a tabled predicate. Figure~\ref{fig:tablePredicate}
shows that \texttt{table/2} retrieves the \texttt{Table} data structure for the
given \texttt{Wrapper} call pattern. There is one table for every
distinct call pattern encountered so far; if the current call pattern has not
been encountered before, \texttt{get\_table\_for\_variant/2} allocates a fresh
data structure for it.

Then \texttt{table/2} switches on the \texttt{Table}'s status. If the status is
\texttt{complete}, it means that all answers for the \texttt{Wrapper} call
pattern are already available in the table. The call is then answered by
consuming the answers with the
\texttt{get\_answer\_from\_table/2} predicate.

Otherwise, we either start collecting answers (\texttt{run\_leader/3}), or we are already in the process
of collecting answers and simply proceed (\texttt{run\_follower/4}). The call that initiates answer
collection is called the \emph{leader}. A leader is a call to a tabled
predicate that has only non-tabled ancestors in the dynamic call graph. Other
calls to tabled predicates during answer collection are called
\emph{followers}. Every follower has a leader as its ancestor.  The leader and
its followers make up a \emph{scheduling component}. Multiple scheduling components can occur during
program execution.

\begin{example}
Consider the top-level call \texttt{?- p(X,Y).} for our running example. Then
\texttt{p(X,Y)} clearly is the leader of a new scheduling component. The recursive call \texttt{p(X,Z)} in the
first clause constitutes a follower in its scheduling component.
\end{example}

\paragraph{The Leader} \label{subsubsec:leader}

The leader, defined in Figure~\ref{fig:handleLeader}, takes responsibility for
computing all the answers of its scheduling component. To quickly identify whether
there currently is a leader, we use a global non-backtrackable variable.
The predicates
\texttt{exists\_scheduling\_component/0} and
\texttt{create\_scheduling\_component/0} check and set this variable. The predicate
\texttt{unset\_scheduling\_component/0} unsets it.

The job of the leader consists of two tasks: 1) it starts computing the answers of
the scheduling component with \texttt{activate/3}, and 2) it computes the least fixpoint for the
whole scheduling component with \texttt{completion/0}.


\begin{figure}
\begin{minipage}{0.43\textwidth}
\begin{MyFloatingVerbatim}
run_leader(Wrapper,Worker,Table) :-
  create_scheduling_component,
  activate(Wrapper,Worker,Table),
  completion,
  unset_scheduling_component.

\end{MyFloatingVerbatim}
\caption{Handling the leader call.\label{fig:handleLeader}}
\end{minipage}
\enskip 
\begin{minipage}{0.54\textwidth}
\begin{MyFloatingVerbatim}
run_follower(fresh,Wrapper,Worker,Table) :-
  activate(Wrapper,Worker,Table),
  shift(call_info(Wrapper,Table)).

run_follower(active,Wrapper,Worker,Table) :-
  shift(call_info(Wrapper,Table)).
\end{MyFloatingVerbatim}
\caption{Handling a follower call.\label{fig:handleNonLeader}}
\end{minipage}
\end{figure}

\paragraph{Followers} \label{subsubsec:nonleader}

Followers, defined in Figure~\ref{fig:handleNonLeader}, have fewer
responsibilities than the leader.  If the table of the follower is \texttt{fresh}, i.e. it is
the first time the call pattern occurs, then the follower \texttt{activate}s
the answer computation. Subsequently, it yields control with \texttt{shift/1};
this is explained in more detail in the next subsection. If the table is already \texttt{active}ly
collecting answers, the follower immediately yields control.

\subsection{Activation and Delimited Answer Computation}
\label{subsec:activation}

When a call pattern is encountered for the first time, the computation of its
answers is activated with the predicate \texttt{activate/3}. This predicate,
defined in Figure~\ref{fig:activation}, alters the table status from
\texttt{fresh}ly allocated to \texttt{active} and puts the \texttt{Worker} to
work with the auxiliary \texttt{delim/3} predicate. Note that a failure driven loop
is used to backtrack over all the alternatives of \texttt{Worker}.

\begin{figure}[h]
\begin{center}
\begin{minipage}{0.65\textwidth}
\begin{MyFloatingVerbatim}
activate(Wrapper,Worker,Table) :-
  table_set_status(Table,active),
  (
    delim(Wrapper,Worker,Table),
    fail
  ;
    true
  ).
\end{MyFloatingVerbatim}
\end{minipage}
\end{center}
\caption{Activation.\label{fig:activation}}
\end{figure}

\begin{figure}
\begin{center}
\begin{minipage}{0.9\textwidth}
\begin{MyFloatingVerbatim}
delim(Wrapper,Worker,Table) :-
   reset(Worker,Continuation,SourceCall),
   ( Continuation == 0 ->
     store_answer(Table,Wrapper)
   ;
     SourceCall = call_info(_,SourceTable),
     TargetCall = call_info(Wrapper,Table),
     Dependency = dependency(SourceCall,Continuation,TargetCall),
     store_dependency(SourceTable,Dependency)
   ).
\end{MyFloatingVerbatim}
\end{minipage}
\end{center}
\caption{Delimited execution.\label{fig:delimCode}}
\end{figure}

The body of a tabled predicate $p/n$ is actually executed by predicate
\texttt{delim/3}, defined in Figure~\ref{fig:delimCode}. This predicate runs
$p/n$'s \texttt{Worker} in the context of a \texttt{reset/3}. If
the \texttt{Worker} succeeds normally, the answer is added to the table with
\texttt{store\_answer/2}.

However, if the \texttt{Worker} calls a tabled predicate $q/m$ ---
with either the same or a different call pattern as $p/n$ --- then
\texttt{Worker} does not terminate normally. The reason is that the $q/m$ call
is a follower, and \texttt{run\_follower/4} always ends in a \texttt{shift/1}
without producing an answer. Instead the \texttt{Worker}
suspends, capturing the remainder in \texttt{Continuation}.

\begin{example}
Consider the following clause from our running example:
\begin{MyNonFloatingVerbatim}
p_aux(X,Y) :- p(X,Z), e(Z,Y).
\end{MyNonFloatingVerbatim}
The worker \texttt{p\_aux(X,Y)} for the call \texttt{p(X,Y)} immediately suspends at the recursive call
\texttt{p(X,Z)} with \texttt{Continuation = e(Z,Y)}.
\end{example}

Through this suspension, we bypass the regular depth-first execution mechanism of Prolog and avoid
its potential non-termination. We replace the depth-first search by the least
fixpoint computation of the \texttt{completion} phase. For this purpose, we
record the suspended computation in the form of a \texttt{dependency/3}
structure. This structure expresses that given an answer for the $q/m$ call, one
may obtain answers for the $p/n$ call by resuming the suspended continuation.
We name $q/m$ the \emph{source call} and $p/n$ the
\emph{target call}. For the source call, it is sufficient to hold on to the
\texttt{SourceTable} to be able to retrieve an answer later. For the target
call, we need the \texttt{Wrapper} in addition to the table, as the
\texttt{Wrapper} contains the partial answer that the continuation will
instantiate. This explains the form of the \texttt{dependency/3} structure,
which is stored in the table of the source call to be
triggered whenever a new answer is added.

\begin{example}
The dependency for our example above expresses that, given an answer for \texttt{p(X,Z)},
we may obtain answers for \texttt{p(X,Y)} by executing \texttt{e(Z,Y)}. For instance,
if we get the answer \texttt{X = a, Z = b} for \texttt{p(X,Z)}, and we have the fact \texttt{e(b,c)}
then we obtain the answer \texttt{X = a, Y = c} for \texttt{p(X,Y)}.
\end{example}

\begin{example}\label{ex:4}
Assume that \texttt{e/2} is defined by the facts \texttt{e(a,b)} and
\texttt{e(b,c)}.  Then the query \texttt{?- p(X,Y)} yields not only the
dependency on \texttt{p(X,Z)} through the first clause of \texttt{p\_aux/2} but
also the answers \texttt{p(a,b)} and \texttt{p(b,c)} through the second clause
of \texttt{p\_aux/2}.  Since \texttt{p(X,Z)} is a variant of \texttt{p(X,Y)}, the dependency and the two answers
are all associated with the same table.
\end{example}

\subsection{Completion}\label{sec:completion}

\begin{figure}[b]
\begin{minipage}{0.99\textwidth}
\begin{MyFloatingVerbatim}
completion :-                  completion_step(SourceTable) :-
  ( worklist_empty ->            (
      set_all_complete,            table_get_work(SourceTable,Answer,
      cleanup_tables                 dependency(Source,Continuation,Target)),
  ;                                Source = call_info(Answer,_),
      pop_worklist(Table),         Target = call_info(Wrapper,TargetTable),
      completion_step(Table),      delim(Wrapper,Continuation,TargetTable),
      completion                   fail
  ).                             ;
                                   true
                                 ).
\end{MyFloatingVerbatim}
\end{minipage}
\caption{The completion fixpoint.\label{fig:fixpoint}}
\end{figure}

The completion phase, defined in Figure~\ref{fig:fixpoint}, computes the
fixpoint over all answers and dependencies of the scheduling component. Just like Datalog's
semi-naive approach~\cite{datalogNeverAsk}, our implementation tries to avoid unnecessary
recomputation. More code details are available in~Appendix~C.

We maintain a worklist of all tables for which at least one associated
answer has not been fed into at least one associated dependency. This worklist
is updated whenever a new answer or new dependency is associated with a table.

Predicate \texttt{completion/0} is the driving loop of the completion phase.
It repeatedly pops a table from the worklist and calls \texttt{completion\_step/1} to process
answer/dependency pairs that have not yet been combined.
When the worklist is empty, the completion fixpoint has been reached. Then \texttt{set\_all\_complete/0} sets the status of every table in the scheduling component to \texttt{complete}. Finally \texttt{cleanup\_tables/0} erases all the dependencies, as they are no longer necessary.

Predicate \texttt{completion\_step/1} retrieves an unprocessed pair \texttt{Answer}/\texttt{Dependency} from
the table by calling \texttt{table\_get\_work/3}. It instantiates the source of
the dependency with the answer and resumes the dependency's continuation with
\texttt{delim/3}, binding the variables in the partial answer \texttt{Wrapper} along the way. This process may lead to new answers or new dependencies that spur
the fixpoint computation on. Here, a failure-driven loop is used to iterate over
all answer/dependency pairs.

\begin{example}
Let us consider the completion that follows Example~\ref{ex:4}. There is one entry in
the worklist: the table for call variant \texttt{p(X,Y)}. This table has two unprocessed
pairs:\footnote{We have abbreviated the call information for the sake of clarity.} 
\[ \begin{array}{c}
    \texttt{p(a,b)}~/~\texttt{dependency(p(X,Z),e(Z,Y),p(X,Y))} \\
    \texttt{p(b,c)}~/~\texttt{dependency(p(X,Z),e(Z,Y),p(X,Y))}
   \end{array}\]
The first pair yields the new answer \texttt{p(a,c)} with the help of the
fact \texttt{e(b,c)}. The second pair yields nothing.
The production of a new answer reschedules the table for \texttt{p(X,Y)} in the worklist. Yet the second
completion round yields no new answers or dependencies and the fixpoint computation
terminates with answer set \{\texttt{p(a,b)}, \texttt{p(b,c)}, \texttt{p(a,c)}\} for call \texttt{p(X,Y)}.
\end{example}
\TIKZ{\tikzset{batch/.style={shape=ellipse,inner sep=0,draw}}}

\TIKZ{\tikzset{dequeue/.style={
  matrix of math nodes,
  column sep=5mm,
  nodes={anchor=base, text height=1mm,text width=3mm,align=center},row sep=0mm}}}

\TIKZ{\tikzset{new/.style={fill=black!10}}}
\TIKZ{\tikzset{old/.style={}}}

\subsection{The Table Data Structures}
\TIKZ{\tikzset{combine/.style={line width=0.3mm,decorate}}}
\TIKZ{\tikzset{emit/.style={color=black!60,-{Stealth[]},decorate,decoration={snake,amplitude=0.5mm,segment length=2mm, post length=2mm},text=black}} \tikzset{dep/.style={-{Stealth[]},dotted,color=black!60,line width=0.2mm}}} 

The central data structure used by the tabling control flow explained above is the \emph{table}. We maintain one such table per call
variant, which can be retrieved from a global repository of all tables.  This
global repository is implemented in the form of a trie data structure, also
known as the call trie, that maps call patterns to tables.

There is a second global data structure, the global worklist, which maintains a
simple queue of tables for the algorithm explained in
the previous subsection.

The table itself consists of two parts: the answer trie and the local worklist:
\begin{itemize}
\item
The answer trie is where \texttt{get\_answer\_from\_table/2}
finds its answers. Moreover, the trie allows \texttt{store\_answer/2} to
quickly check whether a newly produced answer has already been computed before,
and to only store it in case it has not.
\item
The local worklist serves the \texttt{table\_get\_work/3} predicate. It
retrieves pairs of answers and dependencies that have not
been combined before. For this purpose we use a dequeue (i.e., a double-ended
queue) that contains answers and dependencies.

The dequeue maintains the invariant that an answer is to the left of a dependency
if and only if they have not been combined. New answers are added on the
left, because they have not been combined with any dependency yet.
New dependencies are added on the right.

For performance reasons, the dequeue batches consecutive answers into
a single entry on insertion; the same happens to consecutive dependencies. Every batch
contains homogeneous elements (either answers or dependencies) and is
implemented as a list --- the position of the elements in the list is
insignificant. Batches of the same type are not merged if they become adjacent during the combination of answers and dependencies. Doing so would reduce the number of swaps, but at the cost of merging the lists. 

The \texttt{table\_get\_work/3} predicate retrieves a batch of answers
immediately to the left of a batch of dependencies, swaps their positions and yields the elements of
their Cartesian products for processing. 
Dependencies and answers that are created by the combination are also sent to the appropriate tables. A single step of this process is illustrated in Figure~\ref{fig:localWorklist}. The solid arrow denotes the transformation of the local worklist. The wavy line denotes the emission of new answers and dependencies that are generated by the completion step. The answers in the gray ellipse have been added to the local worklist, and will eventually move to the right of all dependencies.

\begin{figure}
\includegraphics{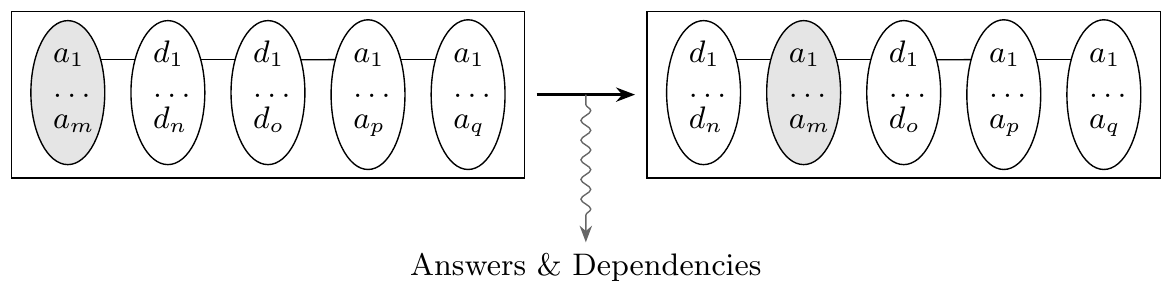}
\caption{Combining answers and dependencies in a local worklist.\label{fig:localWorklist}}
\end{figure}

\end{itemize}

\paragraph{Implementation Support}
The key Prolog implementation support for these tables are mutable terms and
non-backtrackable mutations (Appendix~B). We also use a global variable for the table
repository. These features are widely available. The non-backtrackable
nature is essential to retain the collected answers and
dependencies across disjunctions.

\subsection{Completion of a Double Recursive Call}

\newcommand{\predicate}{r}

\begin{example}
Consider a variant of our running example where the recursive clause is replaced by:
\begin{MyNonFloatingVerbatim}
r(X,Y) :- r(X,Z), r(Z,Y).
\end{MyNonFloatingVerbatim}
Figure~\ref{fig:ex:r} illustrates the computation of \texttt{?- \predicate(a,Y)}. Each table is a rectangle. The consecutive states of its worklist are shown from top to bottom. A dotted arrow shows the target of a dependency. The solid and wavy lines are as in Figure~\ref{fig:localWorklist}. In the explanation, the labels of the completion steps in the figure are written between parentheses.
The call \texttt{?- \predicate(a,Y).} gives rise to the dependency
\texttt{D1 = dependency(\predicate(a,Z),\predicate(Z,Y),\predicate(a,Y))} and the answer \texttt{\predicate(a,b)} (left rectangle).

\paragraph{Iteration 1} In the first iteration of completion (1), the answer is fed into the dependency (wavy arrow $\alpha$), hence \texttt{D1} and \texttt{\predicate(a,b)} are swapped.
This exposes the call \texttt{\predicate(b,Y)} (middle rectangle). For this new call we immediately obtain the
dependency \texttt{D2 = dependency(\predicate(b,Z1),\predicate(Z1,Y),\predicate(b,Y))} and the answer \texttt{\predicate(b,c)}.
We also record dependency \texttt{D3 = dependency(\predicate(b,Y),true,\predicate(a,Y))} between
\texttt{\predicate(b,Y)} and \texttt{\predicate(a,Y)}. The \texttt{true} in \texttt{D3} represents the empty continuation: finding an answer for \texttt{\predicate(b,Y)} gives an answer for \texttt{\predicate(a,Y)} for free!

\TIKZ{\tikzset{table/.style={draw,matrix of nodes,nodes={batch,inner sep=0.5mm,anchor=base,outer sep =0mm, align=center},column sep=0mm,row sep=6mm}}}
\begin{figure}
 \vspace{-5mm}\includegraphics{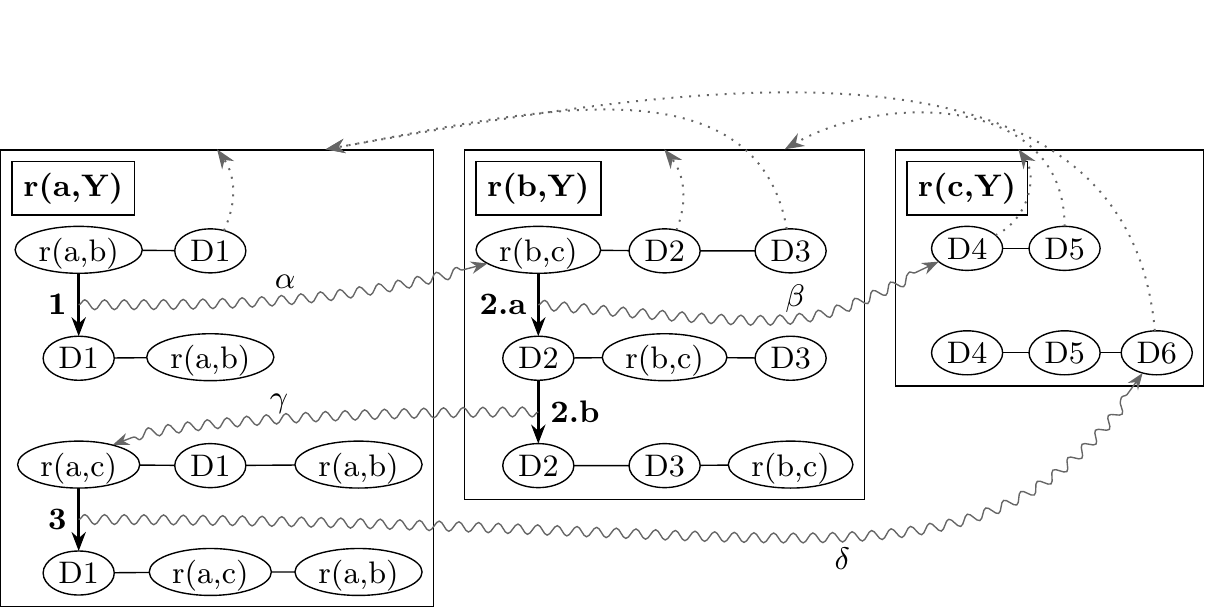}
 \caption{\label{fig:ex:r}Illustration of the computation of \texttt{r(a,Y)}.}
\end{figure}

\paragraph{Iteration 2}
During the second iteration, we feed the answer \texttt{\predicate(b,c)} into the two
dependencies \texttt{D2} (2.a, wavy arrow $\beta$) and \texttt{D3} (2.b, wavy arrow $\gamma$).
\begin{enumerate}
\item[$\beta$]
In the \texttt{D2} case, we expose a new
call \texttt{\predicate(c,Y)} (right rectangle) yielding no direct answer, but a new dependency
\texttt{D4 = dependency(\predicate(c,Z2),\predicate(Z2,Y),\predicate(c,Y))} and a derived dependency
\texttt{D5 = dependency(\predicate(c,Y),true,\predicate(b,Y))}.
\item[$\gamma$]
In the \texttt{D3} case, we obtain the new answer \texttt{\predicate(a,c)} for the top-level call.
\end{enumerate}

\paragraph{Iteration 3}
During the third iteration (3), we feed the new answer into dependency \texttt{D1} (wavy arrow $\delta$).
This yields the call \texttt{\predicate(c,Y)} and the dependency \texttt{D6 = dependency(\predicate(c,Y),true,\predicate(a,Y))}.

\paragraph{The fixpoint}
Finally, there is no more work to be done: at the bottom of each rectangle, all D$i$ are left of all answers. Hence, the fixpoint comprises the
answer table $\{\texttt{\predicate(a,b)}, \texttt{\predicate(a,c)}\}$ for the call pattern \texttt{\predicate(a,Y)},
the answer table $\{\texttt{\predicate(b,c)}\}$ for the call pattern \texttt{\predicate(b,Y)} and the
empty answer table for \texttt{\predicate(c,Y)}.

\end{example}

%% file: evaluation.tex
\subsection{Implementation Effort}

Table~\ref{tab:codeSize} summarizes the implementation effort in lines of
Prolog (LoC).  The control flow shown in this paper comprises 60 LoC, or
less than 11\% of the overall effort.  The majority goes to the two kinds of
data structures, the tries (40\%) and the worklists (45\%). Adding 25 lines of
glue code, this amounts to an implementation for 577 Prolog LoC.

\begin{table}[b]
\begin{center}
{ \footnotesize
\ra{1.3}
\begin{tabular}{lr@{\qquad}lr}
\toprule
\textbf{Category} & \textbf{LoC} & \textbf{Category} & \textbf{LoC} \\
\midrule
Control flow          & 60  & Completion Worklists & 259 \\
Call and Answer Tries & 233 & Miscellaneous        & 25 \\
\midrule
Total                                  & 577 \\
\bottomrule
\end{tabular}
}
\caption{Code size in lines of code.\label{tab:codeSize}}
\end{center}
\end{table}

\subsection{Performance}

While raw efficiency is not the main objective of our lightweight
implementation, it is nevertheless important to achieve a reasonable
performance compared to the existing state-of-the-art tabling systems.
In order to evaluate this, we compare our implementation in hProlog 3.2.38
against XSB 3.4.0 \cite{swift2012}, B-Prolog 8.1 \cite{bprolog}, Yap 6.3.4 \cite{yap} and Ciao 1.15-2731-g3749edd \cite{ciao} on a
number of benchmarks.\footnote{The description and code of the benchmarks can be found at \url{http://users.ugent.be/~bdsouter/tabling/}.}  Table~\ref{tab:grandTableTimings} summarizes the results
(in ms) obtained on a Dell PowerEdge R410 server (2.4 GHz, 32 GB RAM) running Debian 7.6. In parentheses, we have indicated the maximum resident set size (RSS) in megabytes and the proportion of hProlog to XSB. 

\paragraph{Discussion}

The XSB system is the reference system for tabling; it has invested most time
and resources in the development of its tabling infrastructure. We see that it
is 8 to 38 times faster than our implementation, but
45 to 78 times faster for two outliers (path right last: binary tree 18 and 10k pingpong). It has a maximum RSS that is up to 7 times as large, and 14 times for path double first 500.
In general, standard trie-based structures overload the memory because representation sharing is poor. This has been addressed by Raimundo and Rocha~\shortcite{raimundo11}. 

Since XSB does not support big integers, it was not meaningful to run
the Fibonacci benchmark, recorded as O/F (for overflow).
This is a case in point for wider tabling support in other
systems: often we need both tabling and other non-standard
features.

B-Prolog is only half as fast as XSB on many benchmarks, but is architecturally different: BProlog implements linear tabling and uses hash tables instead of tries.
Moreover, in several
cases B-Prolog is notably slower than XSB (i.e., n-reverse) and even much
slower than our own implementation (recognize, shuttle, ping pong). Yet, unlike
XSB, B-Prolog does support big integers and is substantially faster than
our approach for the fib benchmark.
All in all the results are mixed and point out several weaknesses in the B-Prolog
implementation compared to our all Prolog implementation.


\newcommand{\millis}[1]{#1}


\newcommand{\megabyte}[1]{ (#1)}

\begin{table}[t]
\begin{center}
\begin{minipage}{\linewidth}
{ \footnotesize
\ra{1.3}
\begin{tabular}{l@{}rr@{}r@{}r@{}r@{}r}
\toprule
\textbf{Benchmark} &\textbf{Size}  &  \multicolumn{1}{l}{\textbf{hProlog}} & $\frac{\mathbf{hProlog}}{\mathbf{XSB}}$  & $\frac{\mathbf{hProlog}}{\mathbf{B-Prolog}}$ & $\frac{\mathbf{hProlog}}{\mathbf{Yap}}$  & $\frac{\mathbf{hProlog}}{\mathbf{Ciao}}$\\
\midrule
\multicolumn{1}{l}{\textbf{fib$^{a}$}}
  & 500    & \millis{24}  \megabyte{13} & O/F (---) & $\infty$ & $\infty$  & --- \\
  & 750    & \millis{33}  \megabyte{13} & O/F (---) & 17      & 41    & --- \\
  & 1,000  & \millis{46}  \megabyte{13} & O/F (---) & 46      & 19    & --- \\
  & 10,000 & \millis{982} \megabyte{66} & O/F (---) & 3       & 44    & --- \\
\multicolumn{1}{l}{\textbf{recognize$^{a}$}} &
  20,000        & \millis{205} \megabyte{73}   & 26 (1) &  0.003  & 11 & 4 \\
& 50,000        & \millis{503} \megabyte{221}   & 30 (2) &  0.001  & 14 & 4 \\
\multicolumn{1}{l}{\textbf{n-reverse$^{a}$}} &
     500     & \millis{767}  \megabyte{138}  & 38  (5)   & 11     & 15  & 45 \\
 & 1,000    & \millis{2,800} \megabyte{537} & 31  (6)    & 6      & 8   & 34 \\
\multicolumn{1}{l}{\textbf{shuttle$^{b}$}} &
   2,000  & \millis{44}    \megabyte{12} & $\infty$ (2) & 0.1  & $\infty$ & 9 \\
 & 5,000  & \millis{138}   \megabyte{14} & 23       (2) & 0.08 & $\infty$ & 12 \\
 & 20,000 & \millis{582}   \megabyte{29} & 24       (4) & 0.02 & $\infty$ & 10 \\
 & 50,000 & \millis{1,586} \megabyte{72} & 29       (6) & 0.01 & $\infty$ & 12 \\
\multicolumn{1}{l}{\textbf{ping pong}} &
     10,000 & \millis{271} \megabyte{16} & 45 (2) & 0.07 & $\infty$ & 14 \\
   & 20,000 & \millis{490} \megabyte{28} & 35 (4) & 0.03 & $\infty$ & 8 \\
\multicolumn{1}{p{2.5cm}}{\textbf{path double first loop}} & 
  50  & \millis{653}   \megabyte{14} & 19 (2) & 13 & $\infty$ & 7 \\
& 100 & \millis{4,638} \megabyte{29} & 17 (4) & 10 & $\infty$ & 6 \\
\multicolumn{1}{l}{\textbf{path double first}} & 
     50  & \millis{162}     \megabyte{12}  & 27 (2) & 15 & $\infty$ & 14 \\
   & 100 & \millis{989}     \megabyte{16}  & 20 (3) & 12 & $\infty$ & 10 \\
   & 200 & \millis{6,785}   \megabyte{53}  & 18 (7) & 16 & $\infty$ & 10 \\
   & 500 & \millis{110,463} \megabyte{267} & 25 (14) & 19 & $\infty$ & 14 \\
\multicolumn{1}{p{2.5cm}}{\textbf{path right last: pyramid 500}} & 
  500 & \millis{1,914}  \megabyte{104} & 35   (7)   & 29   & $\infty$ & 27 \\
\multicolumn{1}{p{2.5cm}}{\textbf{path right last: binary tree 18}} & 
18 & \millis{108,662} \megabyte{4,120} & 78 (5) & 50 & 3,461 & 42 \\
\multicolumn{1}{l}{\textbf{test large joins 2$^{c}$}} & 
  12 & \millis{3,001} \megabyte{237} & 10 (5)   & 4    & $\infty$ & 12 \\
\multicolumn{1}{l}{\textbf{joins mondial}} & 
  & \millis{6,444} \megabyte{399} & 8 (2) & 7 & 224 & 6 \\
\bottomrule\\
	\end{tabular}
	}

	\caption{Results of the performance benchmarks.\label{tab:grandTableTimings}}
	\smallskip
	{ \footnotesize
  Source:
  \textit{a} \cite{fan1992} \quad
  \textit{b} \cite{cat1998} \quad
  \textit{c} Yap benchmark suite
  } 
	\end{minipage}
	\end{center}
	\end{table}

The Yap tabling implementation, which is based on that of XSB, is clearly the
fastest: the underlying engine is much faster~\cite{yaptab2000}. It outperforms our approach on all benchmarks, and the other
systems on most. Many benchmarks take less than \unit{1}{\milli\second}, rounded down to \unit{0}{\milli\second}, hence the factor $\infty$ in the table.

The performance of Ciao lies between that of XSB and B-Prolog. Performance of our implementation is within a factor 4 to 14 of Ciao, with reverse and path right last as outliers. Running the Fibonacci benchmarks is currently not possible, as tabling and bignums currently do not operate together\footnote{Personal email communication with Manuel Carro.}.

\paragraph{Summary}
We consider the performance results of our implementation very reasonable,
especially if we take into account the stark contrast between our lightweight
pure Prolog implementation and the complex integration in other systems.
As part of future work, we think that advances in three areas may
positively affect performance.
Firstly, continuations are copied with \texttt{copy\_term/2}. A special-purpose \texttt{copy\_continuation/2} could do better by exploiting the known structure of these terms. Other applications using delimited control could
benefit from this optimization as well.
Secondly, we don't statically identify strongly connected components in the scheduling component.
Doing so would allow the specialisation of completion.
Finally, in contrast with state-of-the art implementations, our tries do not use
substitution factoring.

%% file: relatedWork.tex
\paragraph{Delimited Control}

While delimited control is well-known in the functional programming world, it
has not received much attention in the context of Prolog. Only recently have
Schrijvers \textit{et al.} provided an unobtrusive implementation in the WAM
~\cite{iclp2013,reset}. In the continuation-passing
implementation~\cite{td94:LOPSTR} of
BinProlog~\cite{DBLP:journals/tplp/Tarau12} this is even easier. Schrijvers
\textit{et al.} also illustrate the power of delimited control by porting
various effect handlers~\cite{plotkinPretnar} to Prolog. As
far as we know, this paper shows the first Prolog-specific
application.

\paragraph{XSB}
XSB is the best-known Prolog engine supporting tabling. Its foundation, SLG resolution, has been described by Chen and Warren~\shortcite{chen1996}. Swift and Warren~\shortcite{swift2012} provide a recent survey. Implementing XSB has required nontrivial changes to the architecture of the WAM. XSB maintains a forest of SLD-trees for a tabled predicate. During the computation, the stack may be frozen several times.

\paragraph{CAT and CHAT}
The CAT is an alternative to the SLG-WAM used in XSB~\cite{cat1998}. Rather than freezing memory areas, CAT uses incremental copies to preserve the execution state of suspended computations. CAT's advantage is that the speed of the underlying abstract machine is not affected for non-tabled execution.
CHAT is an improved scheme incorporating some ideas from the SLG-WAM~\cite{chat1998}. CAT and CHAT do require changes to the WAM, but acknowledge that the complexity and scope of these changes should be kept limited.

\paragraph{Linear Tabling}
Linear tabling mechanisms \cite{zhou2000}, which implement the SLDT-resolution strategy, maintain a single execution tree, hence there is the need to steal choicepoints from a former variant call. Each tabled call can be both a producer and a consumer. Similar to our approach, there is no overhead for standard SLD-resolution, but the need for recomputation of subgoals cannot always be avoided. Although simpler than SLG resolution, implementing SLDT still requires the addition of 4 new specifically designed WAM-instructions, a new frame structure and a new data area. Unlike for suspension-based mechanisms, the cut operator works for a class of useful programs.

\paragraph{DRA}
The DRA \cite{guo2001,guo2004} has a goal similar to our approach. The technique implements tabled evaluation without stack-freezing. It postpones clauses containing variant calls at runtime, which is similar to our suspension creation. But to implement this technique, Guo and Gupta introduced six new WAM instructions. 
Compared to XSB, Guo and Gupta's implementation of DRA has a significantly better space performance
, but a worse time performance. The authors cite as sources for XSB's better time performance that XSB avoids reconstructing the execution environment for applying looping alternatives, and secondly that XSB includes tabling in the compiling stage. Both reasons are equally applicable to our approach.

%% file: conclusion.tex
In order to enable a more widespread adoption of tabling, we have presented a
lightweight implementation of tabling on top of delimited control. In contrast
to existing approaches, our approach is implemented entirely in Prolog and
requires no deep modifications to the WAM or complex program transformations.
While there is obviously a trade-off between the simplicity of the
implementation and runtime performance, we believe that the current performance of our
approach is reasonable. Of course, there is ample opportunity for improvement.

In the future we would also like to extend our approach with mode-directed
tabling~\cite{mode,mode:yap}. Our initial exploration has shown that this would only
require a small change to the trie structure.